\documentclass[11pt]{article}
\usepackage[dvips]{graphicx}
\usepackage{mathrsfs}
\usepackage{eucal}
\usepackage{amsmath,amsthm,amssymb}
\usepackage{wrapfig}
\usepackage[dvips]{color}
\usepackage{cite}
\usepackage{framed}
\usepackage{tabularx}
\usepackage[sectionbib]{chapterbib}
\usepackage{threeparttable}
\usepackage{float}
\definecolor{shadecolor}{gray}{0.80}
\DeclareMathVersion{chem}
\SetSymbolFont{letters}{chem}{OT1}{cmr}{m}{n}
\definecolor{shadecolor}{rgb}{0.9, 0.9, 0.90}

\newcommand{\CF}{C_{\textit{\textsf{\hspace{-0.3mm}F}}}}
\newcommand{\NA}{N_{\textit{\textsf{\hspace{-0.3mm}Av}}}}

\newcommand{\G}{\textit{\textsf{G}}}
\newcommand{\HG}{\hspace{1.3mm}\Hat{\textit{\textsf{\hspace{-1.3mm}G}}}\hspace{0.5mm}}

\begin{document}
\renewcommand{\figurename}{\small{Fig.}~}
\renewcommand{\thefootnote}{$\dagger$\arabic{footnote}}

\begin{flushright}
\textit{Excluded Volume Effects of Branched Molecules}
\end{flushright}
\vspace{1mm}

\begin{center}
\setlength{\baselineskip}{25pt}{\LARGE\textbf{Volume Expansion of Branched Polymers}}
\end{center}
\vspace*{5mm}

\vspace*{0mm}
\begin{center}
\large{Kazumi Suematsu} \vspace*{2mm}\\
\normalsize{\setlength{\baselineskip}{12pt} 
Institute of Mathematical Science\\
Ohkadai 2-31-9, Yokkaichi, Mie 512-1216, JAPAN\\
E-Mail: suematsu@m3.cty-net.ne.jp,  Tel/Fax: +81 (0) 593 26 8052}\\[8mm]
\end{center}

\hrule
\vspace{0mm}
\begin{flushleft}
\textbf{\large Abstract}
\end{flushleft}
The excluded volume effects of randomly branched polymers are investigated. To approach this problem we assume the Gaussian distribution of segments around the center of gravity. Once this approximation is introduced, we can make use of the same method as employed for linear molecules. By simulating a model-polymer system, it is found that the excluded volume effects of branched polymers are manifested pronouncedly under any conditions from the dilution limit to the melt, including the $\Theta$ state; every result satisfies the restraining condition: $\left\langle s^{2}\right\rangle^{1/2}\ge N^{1/d}$ in accord with our experiences. As a result the Gaussian approximation extracts the essential features of the excluded volume effects of branched molecules.\\[-3mm]
\begin{flushleft}
\textbf{\textbf{Key Words}}:
\normalsize{Branched Molecules/ Excluded Volume Effects/ Gaussian Approximation/ Inhomogeneity Term/ Concentration Dependence}\\[3mm]
\end{flushleft}
\hrule
\vspace{3mm}
\setlength{\baselineskip}{13pt}
\section{Introduction}
The excluded volume problem of branched molecules is the last frontier in polymer physics. To date, only a few experimental and theoretical works have been reported on this problem\cite{Issacson8, Lubensky8, Redner8, Parisi8, Seitz8, deGennes8, Daoud82, Rosa8, Iwamoto8}. Branched molecules can be produced by means of polyaddition/polycondensation of multifunctional monomers, cross-linking of linear molecules and so forth. Chemical and physical structures of resultant branched molecules are interpreted on the basis of the assumption that the molecules are made randomly under the principle of the equal reactivity of functional units. In this paper, we investigate the problem of the excluded volume effects of branched polymers synthesized according to this random reaction, but without ring closure. To develop the theoretical analysis, it is essential to obtain some information about the distribution of segments. Unfortunately, to date little is known about the segment distribution function of a branched molecule even for the unperturbed one. For this reason we introduce the assumption of the Gaussian distribution. This is a strong assumption\cite{Burchard8}. Then the present work is necessarily focused on the problem of whether a theory constructed on this assumption extracts the core features of the excluded volume effects of branched polymers.

\section{Local Free Energy}
The local free energy of mixing polymers and solvents has the form:
\begin{equation}
\delta G_{mixing}=\delta H_{mixing}-T\delta S_{mixing}\label{BEV-1}
\end{equation}
So we can discuss the enthalpy and the entropy terms separately. We solve eq. (\ref{BEV-1}) according to the lattice model\cite{Flory8}.  We directly follow the standard lattice representation, and assume the random occupation of sites.

\subsection{Mixing Entropy}

\begin{wrapfigure}{r}{7cm}
\vspace{-5mm}
\begin{center}
\includegraphics[width=6cm]{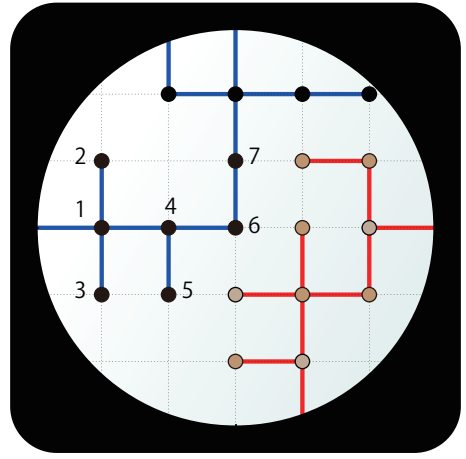}
\caption{Overlapping of segments of branched molecules in the small volume element $\delta V$.}\label{B-LocalFreeEnergy}
\end{center}
\end{wrapfigure}

Consider the small volume element $\delta V$ comprising $\delta m_{0}$ cells which can be occupied by $\delta n_{1}$ solvent molecules and $\delta b$ segments of branched molecules, so that $\delta m_{0}=\delta b+\delta n_{1}$. We identify a segment with a local section (including branching units) on a polymer molecule so that the segment volume is equal to the solvent volume. Suppose that each segment has $z$ neighboring sites. Let those segments stem from some different molecules. Every segment in the volume element $\delta V$ is indexed, in order, such that $k=1, 2, \cdots$. By the assumption, each site is randomly occupied by a segment or a solvent molecule. Let $g_{k}$ be the probability that a site is occupied by a segment when $k$ segments are already put in the volume element $\delta V$, so that $g_{k}=k/\delta m_{0}$. The mean number of the neighboring sites which are vacant is $\sum_{m=0}^{z_{k}-r_{k}}m\binom{z_{k}-r_{k}}{m}\left(1-g_{k}\right)^mg_k^{z_{k}-r_{k}-m}=(z_{k}-r_{k})(1-g_{k})$, where $r_{k}$ is the number of sites forbidden because of the presence of the neighboring segments contiguous with a given segment; this number is 1 for chain molecules, while $r_{k}\ge1$ for branched molecules because of the presence of the branching unit. For the model polymer system illustrated in Fig. \ref{B-LocalFreeEnergy}, $r_{k}=1$ for $k=1$ and 2, $r_{k}=2$ for $k=3$ and $r_{k}=3$ for $k=4$, and so forth. There is a finite probability that a given segment overlaps with the other segments on the same molecule, so that the number of vacant sites for the segment is, depending on the location, $k$, always smaller than is expected from the pseudo self-avoiding molecule assumption\cite{Kazumi82}. Hence $z$ must be replaced with $z_{k}$ ($z_{k}\le z\,$). To date $z_{k}$'s are unknown\cite{Parisi8, deCloizeaux8, Plischke8}, but it has no effects on our final result, as will be seen shortly. Then the total configurations that the $\delta b$ segments can take\,\footnote{\,The formulation of eq. (\ref{BEV-2}) incorporates the segment connectivity only partially\cite{Saito8}. However, if both $k$ and $\delta m_{0}$ are very large numbers (this is true for the real system!), the approximation will be sufficiently exact.} will be
\begin{equation}
\Omega_{\delta V}=\prod_{k=0}^{\delta b-1}(z_{k}-r_{k})(1-g_{k})\label{BEV-2}
\end{equation}
Eq. (\ref{BEV-2}) represents the total number of self-avoiding configurations. The local entropy is then of the form:
\begin{equation}
\delta S_{mixing}+\delta S_{0}=k\log\Omega_{\delta V}=k\left\{\sum_{k=0}^{\delta b-1}\log\, (z_{k}-r_{k})+\log\frac{\delta m_{0}!}{\delta m_{0}^{\delta b}(\delta m_{0}-\delta b)!}\right\}\label{BEV-3}
\end{equation}
Using the Stirring formula, eq. (\ref{BEV-3}) can be approximated as
\begin{equation}
\delta S_{mixing}+\delta S_{0}=-k\left\{\delta n_{1}\log v_{1}+\delta b-\sum_{k=0}^{\delta b-1}\log\, (z_{k}-r_{k})\right\}\label{BEV-4}
\end{equation}
where $v_{1}=\delta n_{1}/\delta m_{0}$ is the volume fraction of the solvent in the volume element, $\delta V$. Since $\delta S_{0}=\delta S_{01}(\delta b=0)+\delta S_{02}(\delta n_{1}=0)=-k\left\{\delta b-\sum_{k=0}^{\delta b-1}\log\, (z_{k}-r_{k})\right\}$, we have the mixing entropy of the pure self-avoiding branched molecules and the pure solvent:
\begin{equation}
\delta S_{mixing}=-k\,\delta n_{1}\log v_{1}\label{BEV-5}
\end{equation}
which is the same form as the expression for the chain molecules\cite{Flory8}.

\subsection{Mixing Enthalpy}
Let $w$ denote the heat of formation for contact. Heat-gain per one contact between a solvent molecule and a segment on a branched polymer may be written as
\begin{equation}
\Delta w=w_{12}-\tfrac{1}{2}(w_{11}+w_{22})\label{BEV-6}
\end{equation}
where the subscripts 1 and 2 denote solvents and polymers, respectively. There are $\delta b$ segments in the volume element $\delta V$, so that there are $(z-\bar{s}_{k})\delta b$ possible sites of such contacts around the $\delta b$ segments, with $\bar{s}_{k}$ being the average segment number of the immediate neighbors. Let us assume that the probability of the contact is proportional to the volume fractions of respective components. Since the solvent volume fraction is $v_{1}$, the heat of formation through the mixing process becomes
\begin{equation}
\delta H_{mixing}=(z-\bar{s}_{k})\Delta w\delta b v_{1}\label{BEV-7}
\end{equation}
The volume fractions of solvent molecules and polymer segments are given, respectively, by
\begin{equation}
\begin{split}
v_{1}=\frac{\delta n_{1}}{\delta n_{1}+\delta b}\\
v_{2}=\frac{\delta b}{\delta n_{1}+\delta b}
\end{split}\label{BEV-8}
\end{equation}
so that $v_{1}=1-v_{2}$. Then eq. (\ref{BEV-7}) may be recast in the form:
\begin{equation}
\delta H_{mixing}=(z-\bar{s}_{k})\Delta w\delta n_{1}v_{2}\label{BEV-9}
\end{equation}
which may further be recast in the known form\cite{Flory8}:
\begin{equation}
\delta H_{mixing}=kT\chi\delta n_{1}v_{2}\label{BEV-10}
\end{equation}
where
\begin{equation}
\chi=\frac{(z-\bar{s}_{k})\Delta w}{kT}\label{BEV-11}
\end{equation}
is the enthalpy parameter that measures the strength of the interaction between a segment and solvent molecules. Only one difference from the linear case is the quantity, $\bar{s}_{k}$.

\subsection{Mixing Free Energy}\label{Mixing Free Energy-B}
Substituting eqs. (\ref{BEV-5}) and (\ref{BEV-10}) into eq. (\ref{BEV-1}), we arrive at the expression for the local free energy in the volume element $\delta V$
\begin{equation}
\delta G_{mixing}=\,kT\left\{\log\,\left(1-v_{2}\right)+\chi v_{2}\right\}\delta n_{1}\label{BEV-12}
\end{equation}
Eq. (\ref{BEV-12}) represents the free energy difference between the mixture of the self-avoiding branched molecules and solvents, and the respective pure components. Eq. (\ref{BEV-12}) is the same formula as derived for linear molecules. It is again realized that the classic theory\cite{Flory8} is constructed on great generality and sound physical basis\cite{Kazumi82}.

In eq. (\ref{BEV-12}), if the parameter $\chi$ can be determined independently, all the constraints of the lattice representation can be removed. Then we can introduce the new definition that the size of the ``segment'' is equal to that of the repeating unit of a polymer molecule.

\section{Expansion Factor}
The expansion factor, $\alpha$, is determined by the force balance between the osmotic potential and the elastic potential. Our final goal is to find out the minimum point of the free energy:
\begin{equation}
\left(\frac{\partial\Delta G}{\partial \alpha}\right)_{T, P}=\left(\frac{\partial\Delta G_{osmotic}}{\partial \alpha}\right)_{T, P}+\left(\frac{\partial\Delta G_{elastic}}{\partial \alpha}\right)_{T, P}=0\label{BEV-13}
\end{equation}

\subsection{Osmotic Potential ($\Delta G_{osmotic}$)}
Let $V_{1}$ be the volume of the solvent. With $\delta n_{1}=(1-v_{2})\delta V/V_{1}$, the free energy of mixing the self-avoiding branched molecules and the solvent becomes
\begin{equation}
\Delta G_{mixing}=\frac{kT}{V_{1}}\int\left(1-v_{2}\right)\left\{\log\,\left(1-v_{2}\right)+\chi v_{2}\right\}\delta V\label{BEV-14}
\end{equation}
The most central point of the theory is to formulate the potential difference between the inside (\textit{hill}: the concentrated region) of a branched molecule and the outside (\textit{valley}: the dilute region). Expanding eq. (\ref{BEV-14}) into the Taylor series and taking the difference, we have
\begin{align}
\Delta G_{osmotic}=&\,\Delta G_{mixing,hill}-\Delta G_{mixing,valley}\notag\\
=&\,\frac{kT}{V_{1}}\int\left\{-\left(1-\chi\right)\mathscr{J}_{1}+\left(1/2-\chi\right)\mathscr{J}_{2}+\frac{1}{6}\mathscr{J}_{3}+\cdots\right\}\delta V\label{BEV-15}
\end{align}
where $\mathscr{J}_{k}'s\, (=v_{hill}^{k}-v_{valley}^{k})$ represent inhomogeneity terms with $v$ denoting the local volume fraction of segments. Eq. (\ref{BEV-15}) is the general expression of the osmotic potential. Because of the constraint, $0\le v_{2}\le 1$, we have collected, for linear molecules, the first two terms only. However, in branched molecules the segments are highly crowded so that the density is much greater in the interior, and we can no longer ignore the higher terms of eq. (\ref{BEV-15}). For this reason we investigate the first three terms.

\subsection{Gaussian Trees}
The excluded volume problem is not rigorously soluble, since the local segment concentration cannot be formulated exactly. In the case of linear molecules, we have assumed the Gaussian distribution. This assumption is not correct. It is well-known that for both the ideal chain and the expanded coil, the segment distribution around the center of gravity is not Gaussian\cite{Ishihara8, Debye8, Mazur8}. Notwithstanding, the consequences derived from the Gaussian approximation are in good accord with experimental observations, and those have given birth to new predictions\cite{Kazumi82}. Encouraged by this success, we apply the same Gaussian distribution approximation to branched molecules.

Before proceeding to theoretical development, it is important to notice that there are two ways to approximate the distribution of expanded molecules. One is the equality employed by Flory\cite{Flory8}:
\begin{equation}
p(x,y,z)dxdydz=\left(\frac{\beta}{\pi\alpha^{2}}\right)^{3/2}\exp\left\{-\beta\left[(x-a)^{2}+(y-b)^{2}+(z-c)^{2}\right]\right\}\alpha^{3}dxdydz\label{BEV-16}
\end{equation}
where $\beta=3/2\langle s_{N}^{2}\rangle_{0}$ with $\langle s_{N}^{2}\rangle_{0}$ being the mean square radius of gyration of an unperturbed molecule; the subscript $N$ denotes the number of segments constituting a branched polymer. The expansion factor, $\alpha$, is defined as $\alpha^{2}=\langle s_{N}^{2}\rangle/\langle s_{N}^{2}\rangle_{0}$. Eq. (\ref{BEV-16}) leads us to the previous expression\cite{Kazumi82} for the segment density:
\begin{equation}
\rho_{N}=N\left(\frac{\beta}{\pi\alpha^{2}}\right)^{3/2}\exp\left\{-\beta\left[(x-a)^{2}+(y-b)^{2}+(z-c)^{2}\right]\right\}\label{BEV-17}
\end{equation}
The other is
\begin{equation}
\Hat{p}(x,y,z)dxdydz=\left(\frac{\beta}{\pi\alpha^{2}}\right)^{3/2}\exp\left\{-\frac{\beta}{\alpha^{2}}\left[(x-a)^{2}+(y-b)^{2}+(z-c)^{2}\right]\right\}dxdydz\label{BEV-18}
\end{equation}
which yields the alternative density formula:
\begin{equation}
\Hat{\rho}_{N}=N\left(\frac{\beta}{\pi\alpha^{2}}\right)^{3/2}\exp\left\{-\frac{\beta}{\alpha^{2}}\left[(x-a)^{2}+(y-b)^{2}+(z-c)^{2}\right]\right\}\label{BEV-19}
\end{equation}
Here $\delta V=\alpha^{3}dxdydz$ for $\rho_{N}$, but $\delta V=dxdydz$ for $\Hat{\rho}_{N}$. Note that eq. (\ref{BEV-16}) is nothing but the distribution function for the unperturbed molecule. In this respect it is obvious that $\Hat{\rho}_{N}$ is an approximation more proper than $\rho_{N}$. Quite unexpectedly when applied to the dilution limit ($C\rightarrow 0$), the two expressions yield, through the triple integral followed by the differentiation, $\partial \iiint \Delta G(\alpha)dxdydz/\partial\alpha$, the same classic result: the fifth power rule, $\alpha^{5}-\alpha^{3}=\text{const.}(1/2-\chi)N^{1/2}$. In contrast, when these formulae are combined with the more general case represented by eq. (\ref{BEV-15}), while $\rho_{N}$ gives the elementary solution, $\Hat{\rho}_{N}$ leads us to complicated mathematics. For this reason we have so far made use of $\rho_{N}$ in place of $\Hat{\rho}_{N}$\cite{Kazumi82}. As is seen shortly, for linear molecules the difference does not arise explicitly between $\rho_{N}$ and $\Hat{\rho}_{N}$, since the inhomogeneity, $\Delta\rho=\rho_{hill}-\rho_{valley}$, disappears very rapidly with increasing $N$ and $\bar{\phi}$. As a result $\rho_{N}$ and $\Hat{\rho}_{N}$ make no appreciable difference, both the expressions reproducing well the experimental points\cite{Daoud81, Westermann8, Graessley8}.

The situation changes radically in branched molecules. Unperturbed branched molecules are unrealistically closely packed, and the segment density increases indefinitely with increasing $N$ as $\rho_{N}\propto N^{\frac{1}{4}}$. Meanwhile it was found that the application of the unperturbed distribution, $\rho_{N}$, leads us to an answer incompatible with the one derived from another viewpoint\cite{Kazumi82}. On this basis in this paper we employ $\Hat{\rho}_{N}$. Then the polymer number concentration in $\delta V$ may be expressed in the form:
\begin{align}
\Hat{C}=&N\left(\frac{\beta}{\pi\alpha^{2}}\right)^{3/2}\sum_{\{a, b, c\}}\exp\left\{-\frac{\beta}{\alpha^{2}}\left[(x-a)^{2}+(y-b)^{2}+(z-c)^{2}\right]\right\}\notag\\
=&N\left(\frac{\beta}{\pi\alpha^{2}}\right)^{3/2}\HG(x, y, z)\label{BEV-20}
\end{align}
Although $\HG$ represents, as with $\G$, a quantity associated with segment concentration at the coordinate $(x, y, z)$, $\HG$ is also a function of $\alpha$. Let $V_{2}$ be the segment volume and we have $\Hat{v}_{2}=V_{2}\Hat{C}$. By virtue of the equality, $dV=d(x-a)d(y-b)d(z-c)=dxdydz$, eq. (\ref{BEV-15}) may be recast in the form:
\begin{equation}
\Delta G_{osmotic}\simeq\frac{kT}{V_{1}}\iiint\left\{-\left(1-\chi\right)\Hat{\mathscr{J}}_{1}+\left(1/2-\chi\right)\Hat{\mathscr{J}}_{2}+\frac{1}{6}\Hat{\mathscr{J}}_{3}\right\}dxdydz\label{BEV-21}
\end{equation}
where $\Hat{\mathscr{J}}_{k}=\Hat{v}_{hill}^{k}-\Hat{v}_{valley}^{k}$. In eq. (\ref{BEV-21}),  we have ignored the higher terms, $\Hat{\mathscr{J}}_{4}, \Hat{\mathscr{J}}_{5}, \cdots$. We are interested in the concentration dependence of the Gibbs potential. To find the equilibrium condition, eq. (\ref{BEV-21}) must be differentiated with respect to $\alpha$ under constant \textit{T} and \textit{P}, namely
\begin{equation}
\left(\frac{\partial\Delta G_{osmotic}}{\partial \alpha}\right)_{T, P}\notag
\end{equation}
Unfortunately the solution is much complicated and appears irreducible to an elementary mathematics. Then our first approximation is to drop the first linear term from the equation so that
\begin{equation}
\left(\partial\iiint\hspace{-1mm}\Hat{\mathscr{J}}_{1}\,dxdydz/\partial \alpha\right)_{T, P}=0\notag
\end{equation}
which amounts to applying $\rho_{N}$ to the first term, but $\Hat{\rho}_{N}$ to the higher terms. Such treatment is not mathematically consistent, but has some physical rationales: firstly, we know that the first term rigorously vanishes at $C=0$; secondly, in terms of the chemical potential, $\Delta\delta\mu$, which is a measure of the osmotic pressure exerted between two imaginary solutions (``\textit{hill}'' and ``\textit{valley}'') separated by a semipermeable membrane, the first term behaves as if a constant, namely, by eq. (\ref{BEV-12}), it vanishes to yield
$$\Delta\delta\mu=\partial\Delta\delta G_{osmotic}/\partial\delta n_{1}=-kT\left\{\left(1/2-\chi\right)(\Hat{v}_{hill}^{2}-\Hat{v}_{valley}^{2})+\frac{1}{3}\left(\Hat{v}_{hill}^{3}-\Hat{v}_{valley}^{3}\right)+\cdots\right\}$$
Thirdly, the excluded volume problem can be interpreted as a phenomenon due to the many-body interaction between segments. On this basis, we introduce the approximation:
\begin{equation}
\left(\frac{\partial\Delta G_{osmotic}}{\partial \alpha}\right)_{T, P}\simeq\frac{kT}{V_{1}}\frac{\partial}{\partial\alpha}\iiint\left\{\left(1/2-\chi\right)\Hat{\mathscr{J}}_{2}+\frac{1}{6}\Hat{\mathscr{J}}_{3}\right\}dxdydz\label{BEV-22}
\end{equation}

A striking aspect of branched molecules distinguished from linear molecules is the fact that unperturbed branched molecules have the extremely compact radius\cite{Zim8, Dobson8, Kajiwara8, Stauffer8, deGennes8}:
\begin{equation}
\begin{aligned}
\langle s_{N}^{2}\rangle_{0}&=\frac{l^{2}}{2N^{2}}\,\frac{\displaystyle N! \{(f-2)N+2\}!}{\displaystyle\{(f-1)N\}!}\sum_{k=1}^{N-1}\binom{(f-1)k}{k-1}\binom{(f-1)(N-k)}{N-k-1}\\
&\simeq \left(\frac{(f-1)\pi}{2^{3}(f-2)}\right)^{1/2} N^{\frac{1}{2}}l^{2}\hspace{5mm}(\text{as}\,\, N\rightarrow \infty)
\end{aligned}\label{BEV-23}
\end{equation}
which shows that the segment density increases indefinitely as $\rho_{N}\approx N/\langle s_{N}^{2}\rangle^{3/2}_{0}\propto N^{1/4}$ for $N\rightarrow\infty$, so it cannot have reality. Because of this anomalous property, the $\Hat{\mathscr{J}}_{3}$ term in eq. (\ref{BEV-22}) (it was negligible in linear molecules) plays an essential role in branched molecules. A natural consequence is that, contrary to the case of linear molecules, the excluded volume effects of branched molecules never disappear under any condition realizable in this real world ($d=3$) including the choice of the solvent-solute combination and the adjustment of the parameter $\chi$. This consequence is necessary in order for the segment density not to diverge at large $N$.

While eqs. (\ref{BEV-19}) and (\ref{BEV-23}) suggest that the density of the unperturbed branched molecule diverges for $N\rightarrow\infty$, it must be that $\Hat{v}_{2}=V_{2}\Hat{C}\le 1$ by definition. This problem can be resolved through the adjustment of the expansion factor, $\alpha$.

\subsection{Elastic Potential ($\Delta G_{elastic}$) for a Single Branched Molecule}
Whether a molecule is linear or branched, the elastic entropy has the form:
\begin{align}
\Delta S=&\,k\log W(\text{deformed})-k\log W(\text{undeformed})\notag\\
=&-\frac{k}{2}\left\{(\alpha_{x}^{2}-1)+(\alpha_{y}^{2}-1)+(\alpha_{z}^{2}-1)\right\}+k\,\log\,(\alpha_{x}\alpha_{y}\alpha_{z})
\end{align}
$S$ is again a function of $\alpha$ alone, and $\alpha_{x}=\alpha_{y}=\alpha_{z}=\alpha$. Hence
\begin{equation}
\Delta S=-\frac{3k}{2}\left(\alpha^{2}-1\right)+3k\,\log\,\alpha
\end{equation}
Since the enthalpy term has already been taken into account in the calculation of the osmotic potential, we have by the thermodynamic relation
\begin{align}
\left(\frac{\partial\Delta G_{elastic}}{\partial \alpha}\right)_{T, P}&=-T\left(\frac{\partial \Delta S}{\partial \alpha}\right)_{T, P} \notag\\
&=\,3kT\left(\alpha -1/\alpha\right)\label{BEV-26}
\end{align}

\subsection{Formulation of Expansion Factor}
The equilibrium expansion factor, $\alpha_{eq}$, can be obtained by substituting eqs. (\ref{BEV-22}) and (\ref{BEV-26}) into eq. (\ref{BEV-13}).
\begin{equation}
\frac{kT}{V_{1}}\frac{\partial}{\partial\alpha}\iiint\left\{\left(1/2-\chi\right)\Hat{\mathscr{J}}_{2}+\frac{1}{6}\Hat{\mathscr{J}}_{3}\right\}dxdydz+3kT\left(\alpha -1/\alpha\right)=0\label{BEV-27}
\end{equation}
It is seen that the expansion factor, $\alpha$, is a strong function of the inhomogeneity, $\Hat{\mathscr{J}}_{k}$, as well as the enthalpy parameter, $\chi$. $\Hat{\mathscr{J}}_{k}$ is the quantity related to the movement of segments from a more concentrated region to a more dilute region, and $\chi$ the quantity closely related to the inflow of solvents into a molecule. It is seen also from eq. (\ref{BEV-27}) that under the $\Theta$ state ($\chi=1/2$), a branched molecule is still in the expanded state because of the existence of the $\Hat{\mathscr{J}}_{3}$ term. Eq. (\ref{BEV-27}) reveals that the molecule should cease to expand as the homogeneous state of $\Hat{\mathscr{J}}_{k}=0$ is approached, which may be interpreted in the following way: Suppose a branched molecule having a large molecular radius $\left\langle s_{N}^{2}\right\rangle$ that can be equated with $\alpha^{2}\left\langle s_{N}^{2}\right\rangle_{0}$, not because of the volume expansion, but because of a large constant term where $\left\langle s_{N}^{2}\right\rangle=\text{constant}\cdot N^{1/2}$. If this polymer satisfies the condition, $\Hat{\mathscr{J}}_{k}=0$, then $\left(\alpha -1/\alpha\right)=0$ by eq. (\ref{BEV-27}), and we have $\alpha=1$. This tells us that when a molecule expands to a large extent so that $\Hat{\mathscr{J}}_{k}\rightarrow0$, it can not expand farther, because the molecule no longer possesses the extra energy to expand beyond that size.

\vspace{3mm}
Unfortunately eq. (\ref{BEV-27}) is generally insoluble analytically except for the limiting case of $C=0$. For this reason we consider the present problem separately for the two cases: the dilution limit and the concentrated solution.

\subsection{Dilution Limit}\label{DilutionLimit}
In the dilution limit ($C\rightarrow 0$) containing a single molecule in the reaction bath, it follows that $\HG_{hill}=\exp\left\{-\frac{\beta}{\alpha^{2}}(x^{2}+y^{2}+z^{2})\right\}$ and $\HG_{valley}=0$. Then integrating the $\Hat{\mathscr{J}}_{k}$ terms in eq. (\ref{BEV-27}) from $-\infty$ to $+\infty$, we obtain
\begin{equation}
\alpha^{5}-\alpha^{3}=N^{2}\frac{V_{2}^{\,2}}{V_{1}}\left(\frac{\beta}{\pi}\right)^{\frac{3}{2}}\left\{\frac{1}{2^{\frac{3}{2}}}\left(1/2-\chi\right)+\frac{V_{2}N}{3^{\frac{3}{2}+1}\alpha^{3}}\left(\frac{\beta}{\pi}\right)^{\frac{3}{2}}\right\}\label{BEV-28}
\end{equation}
Using the generalized expression $\HG_{hill}=\exp\left\{-\frac{\beta}{\alpha^{2}}(x_{1}^{2}+x_{2}^{2}+ \cdots +x_{d}^{2})\right\}$, we may recast eq. (\ref{BEV-28}) in the form:
\begin{equation}
\alpha^{d+2}-\alpha^{d}=N^{2}\frac{V_{2}^{\,2}}{V_{1}}\left(\frac{\beta}{\pi}\right)^{\frac{d}{2}}\left\{\frac{1}{2^{\frac{d}{2}}}\left(1/2-\chi\right)+\frac{V_{2}N}{3^{\frac{d}{2}+1}\alpha^{d}}\left(\frac{\beta}{\pi}\right)^{\frac{d}{2}}\right\}\label{BEV-29}
\end{equation}
with $\beta=d/2\langle s_{N}^{2}\rangle_{0}$.

\subsubsection{Good Solvents}
Let $\langle s_{N}^{2}\rangle^{\frac{1}{2}}\propto N^{\nu}$ for $N\rightarrow\infty$. Then eq. (\ref{BEV-29}) gives for good solvents
\begin{equation}
\langle s_{N}^{2}\rangle=\alpha^{2}\langle s_{N}^{2}\rangle_{0}\propto N^{\frac{4\left(1+\nu_{0}\right)}{d+2}}\label{BEV-30}
\end{equation}
with the subscript 0 denoting the unperturbed state, so that 
\begin{equation}
\nu_{C\rightarrow 0}=\frac{2\left(1+\nu_{0}\right)}{d+2} \label{BEV-31}
\end{equation}
in agreement with the Issacson-Lubensky result\cite{Issacson8, Lubensky8}. Since $\nu_{0}=1/4$ for branched molecules\cite{Zim8, Dobson8, Kajiwara8, Kazumi82}, this gives $\nu_{C\rightarrow 0}=5/8=0.625$ for $d=2$ and $1/2$ for $d=3$, which are commensurate with the simulation result on lattices, 0.615 and 0.46 respectively, by Seitz and Klein\cite{Seitz8}. If we take into account the fact that the simulation has been performed for smaller molecules, $N=20-600$, on the square lattice and, $N=10-60$, on the cubic lattice, the agreement is satisfactory.

\subsubsection{$\Theta$ Solvents}
Substituting $\chi=1/2$ into eq. (\ref{BEV-29}), we have
\begin{equation}
\alpha^{2d+2}-\alpha^{2d}=N^{3}\frac{V_{2}^{\,3}}{3^{\frac{d}{2}+1}V_{1}}\left(\frac{\beta}{\pi}\right)^{d}\label{BEV-32}
\end{equation}
which, for $N\rightarrow\infty$, gives
\begin{equation}
\langle s_{N}^{2}\rangle=\alpha^{2}\langle s_{N}^{2}\rangle_{0}\propto N^{\frac{3+2\nu_{0}}{(d+1)}}\label{BEV-33}
\end{equation}
Hence
\begin{equation}
\nu_{\Theta}=\frac{3+2\nu_{0}}{2(d+1)}\label{BEV-34}
\end{equation}
in agreement with the Daoud result\cite{Daoud82}.

\subsection{Concentrated Solutions}
According to eq. (\ref{BEV-22}), the excluded volume effects in concentrated solutions are dictated by the inhomogeneity terms:
\begin{align}
\iint\cdots\int\hspace{-1mm}\Hat{\mathscr{J}}_{k}\,dx_{1}dx_{2}\cdots dx_{d}=&\left(V_{2}N\right)^{k}\left(\frac{\beta}{\pi\alpha^{2}}\right)^{\hspace{-1mm}\frac{3}{2}k}\iint\cdots\int\left(\HG_{hill}^{\,k}-\HG_{valley}^{\,k}\right)dx_{1}dx_{2}\cdots dx_{d}\notag\\
=&\left(V_{2}N\right)^{k}\left(\frac{\beta}{\pi\alpha^{2}}\right)^{\hspace{-1mm}\frac{3}{2}k}J_{\alpha}^{\,k}\label{BEV-35}
\end{align}
($k=2, 3$), so no simple scaling relations may appear to exist. Let us examine this problem more closely by simulating the hypothetical branched polymer solutions of the R$-$A$_{f}$ model ($f = 3$). As in the case of linear molecules, we solve eq. (\ref{BEV-27}) according to the lattice model ($d=3$): branched polymers are put on the sites of the simple cubic lattice with the unit lengths ($p\times p\times p$), so that
\begin{equation}
\bar{\phi}=V_{2}\frac{N}{p^{3}}\label{BEV-36}
\end{equation}
The inhomogeneity terms are approximated by the integral in the intervals: $[-p/4,p/4]$ for $\HG_{hill}$ for each axis and $[p/4,3p/4]$ for $\HG_{valley}$, so that $J_{\alpha}^{\,k}$ may be recast in the form ($d=3$):
\begin{equation}
J_{\alpha}^{\,k}=\iint\cdots\int_{-p/4}^{p/4}\HG^{\,k}\,dxdydz-\iint\cdots\int_{p/4}^{3p/4}\HG^{\,k}\,dxdydz\label{BEV-37}
\end{equation}
($k=2, 3$). Then eq. (\ref{BEV-27}) can be solved numerically, with the help of eqs. (\ref{BEV-20}), (\ref{BEV-36}) and (\ref{BEV-37}), as functions of the average volume fraction, $\bar{\phi}$, of polymers and the degree of polymerization, $N$.

\section{Assessment of Theory}
Before proceeding with our discussion, it is necessary to have a confirmation that the equation (\ref{BEV-27}) has sound physical basis. For this purpose let us put eq. (\ref{BEV-27}) to the test by applying to the poly(styrene)(PSt, $N$=1096)$-$CS$_{2}$ and the PMMA($N$=5900)$-$CHCl$_{3}$ systems\cite{Cotton8, Daoud81, Westermann8, Graessley8}.
\begin{figure}[h]
\begin{center}
\begin{minipage}[t]{0.46\textwidth}
\begin{center}
\includegraphics[width=7.8cm]{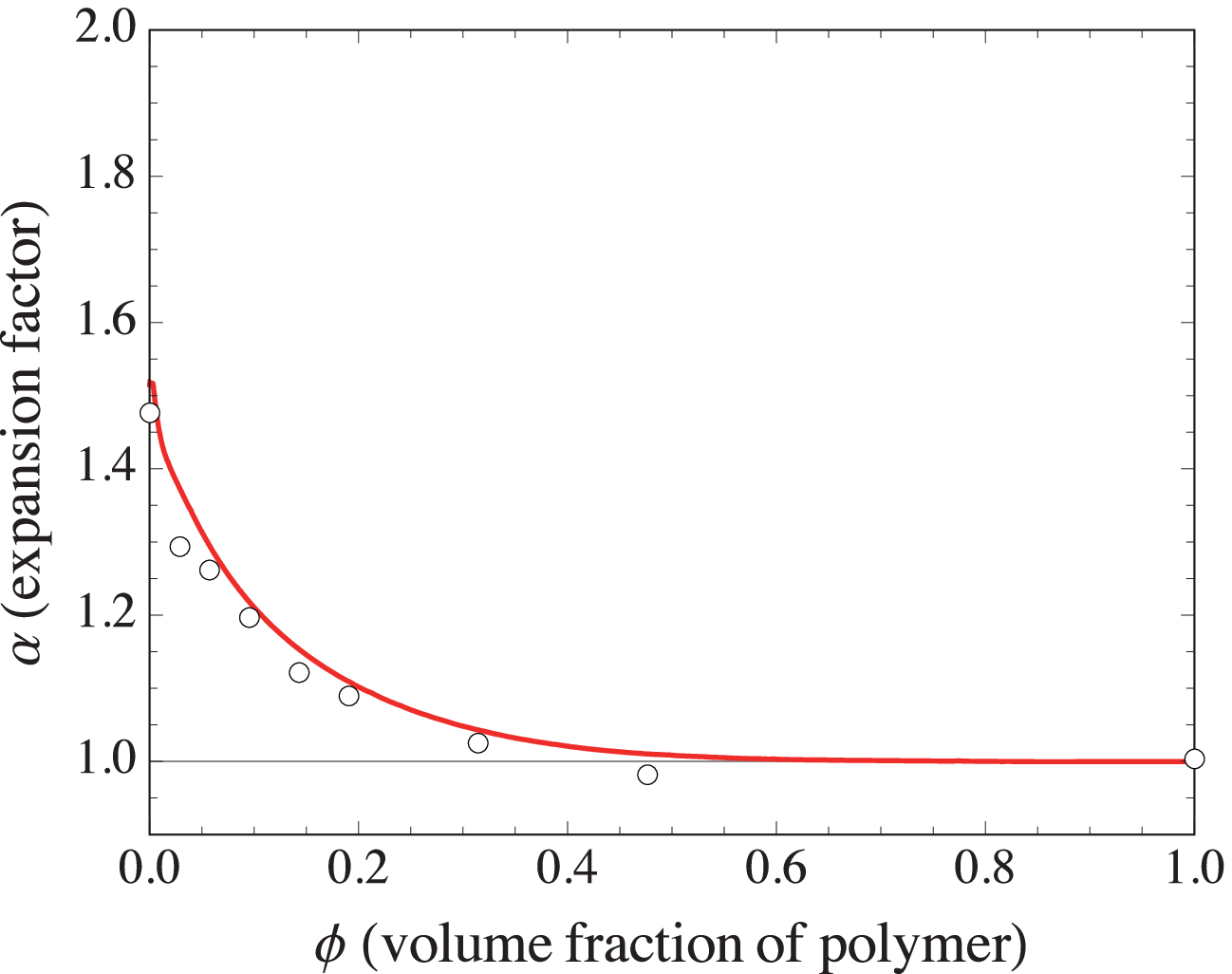}
\end{center}
\vspace{-2mm}
\caption{Expansion factor vs $\bar{\phi}$ plot for PSt$-$CS$_{2}$. Solid line (\textcolor{black}{$-$}): theoretical line by eq. (\ref{BEV-27}) for $\chi=0.4$; open circles ($\circ$): observed points by Daoud and coworkers (the observed value of 82 \text{\AA} at $\bar{\phi}=1$ was replaced by the revised one 93 \text{\AA})\cite{Daoud81}.}\label{PSt-8}
\end{minipage}
\hspace{10mm}
\begin{minipage}[t]{0.46\textwidth}
\begin{center}
\includegraphics[width=7.8cm]{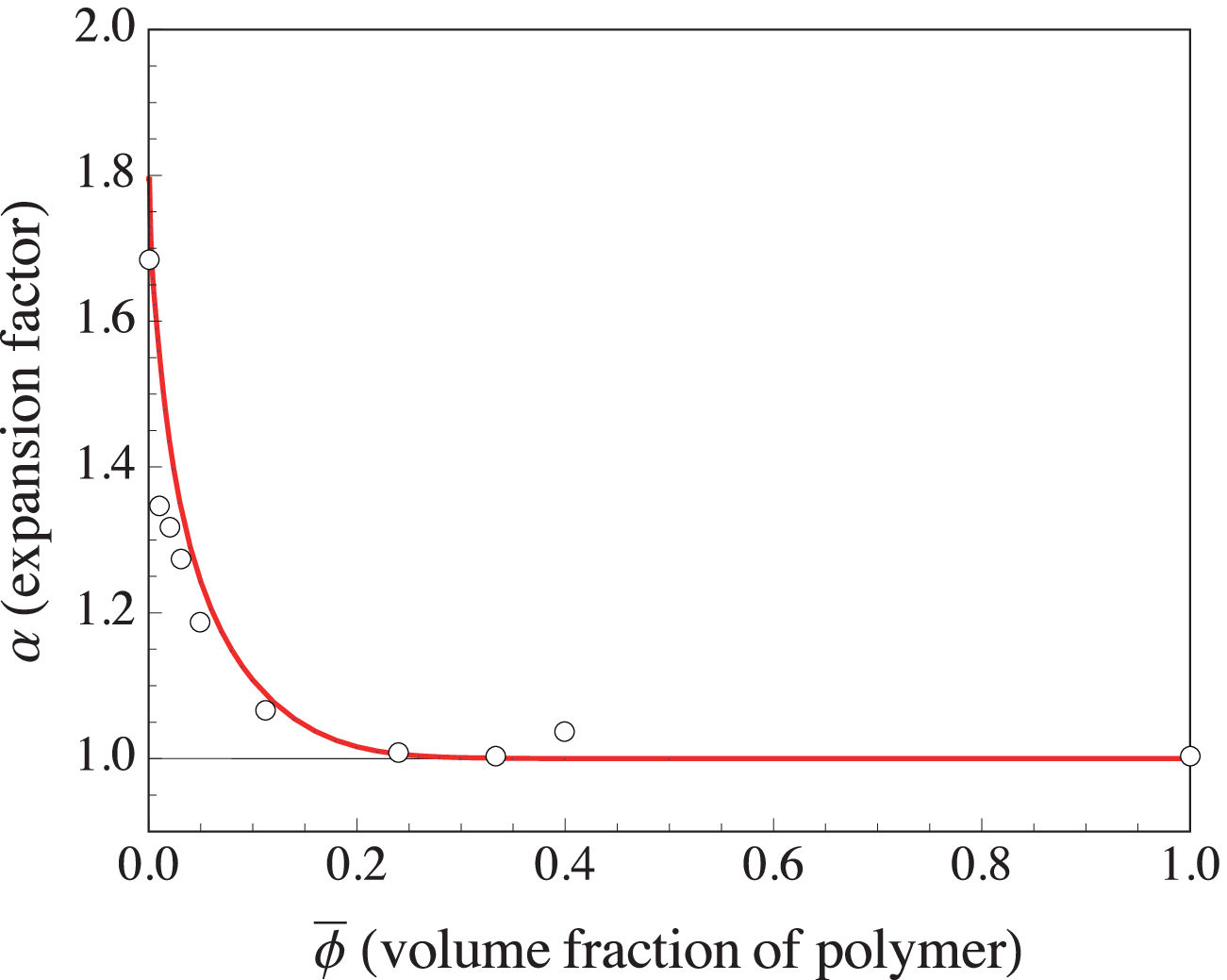}
\end{center}
\vspace{-2mm}
\caption{Expansion factor vs $\bar{\phi}$ plot for PMMA$-$CHCl$_{3}$. Solid line (\textcolor{black}{$-$}): theoretical line by eq. (\ref{BEV-27}) for $\chi=0.3$; open circles ($\circ$): observed points by Cheng, Graessley and Melnichenko\cite{Graessley8}.}\label{PMMA-8}
\end{minipage}
\end{center}
\vspace*{-4mm}
\end{figure}
The simulation was performed using the data employed in the preceding papers\cite{Kazumi82} assuming $\Hat{\mathscr{J}}_{3}=0$. The results are illustrated in Figs. \ref{PSt-8} and \ref{PMMA-8}. Comparing these figures with the corresponding figures in the preceding papers\cite{Kazumi82} which were theorized using $\rho_{N}$ (eq. (\ref{BEV-17})), it is found that $\Hat{\rho}_{N}$ (eq. (\ref{BEV-19})) and $\rho_{N}$ yield almost the same curves. Even though there are some fine differences between them, for instance $\Hat{\rho}_{N}$ yields softer curvatures than $\rho_{N}$, both the equations reproduce well the observed points within the experimental error.

Having confirmed the soundness of the approximate formula (\ref{BEV-27}), let us apply this formula to the excluded volume problem of branched molecules in concentrated region.

\section{Simulation of a Branched Polymer System}
Consider the branched polymer solutions of the R$-$A$_{f}$ model having $f=3$. We give this polymer system the parameters shown in  Table \ref{BPADC-Table} (the mean bond length $\bar{l}$ and the enthalpy parameter $\chi$ are arbitrary). This system is a hypothetical one, but is roughly modeled after the cyclotrimerization polymer of bisphenol A dicyanate in N-methylpyrrolidone solution\cite{Stutz8}.

\begin{center}
  \begin{threeparttable}[h]
    \caption{Parameters of a hypothetical branched polymer solution ($d=3$)}\label{BPADC-Table}
  \begin{tabular}{l l c r}
\hline\\[-1.5mm]
& \hspace{10mm}parameters & notations & values \,\,\,\,\\[2mm]
\hline\\[-1.5mm]
branched polymer & volume of a solvent (NMP\tnote{\,a}\,\,\,) & $V_{1}$ & \hspace{5mm}160 \text{\AA}$^{3}$\\[1.5mm]
& volume of a segment & $V_{2}$ & \hspace{5mm}387 \text{\AA}$^{3}$\\[1.5mm]
& mean bond length & $\bar{l}$ & \hspace{5mm}10 \text{\AA}\,\,\,\\[1.5mm]
& enthalpy parameter & $\chi$ & \hspace{5mm}0 \,\,\,\,\,\,\,\,\\[2mm]
\hline\\[-6mm]
   \end{tabular}
    \vspace*{2mm}
   \begin{tablenotes}
     \item a. N-methylpyrrolidone.
   \end{tablenotes}
  \end{threeparttable}
  \vspace*{4mm}
\end{center}

\begin{figure}[h]
\begin{center}
\includegraphics[width=9.2cm]{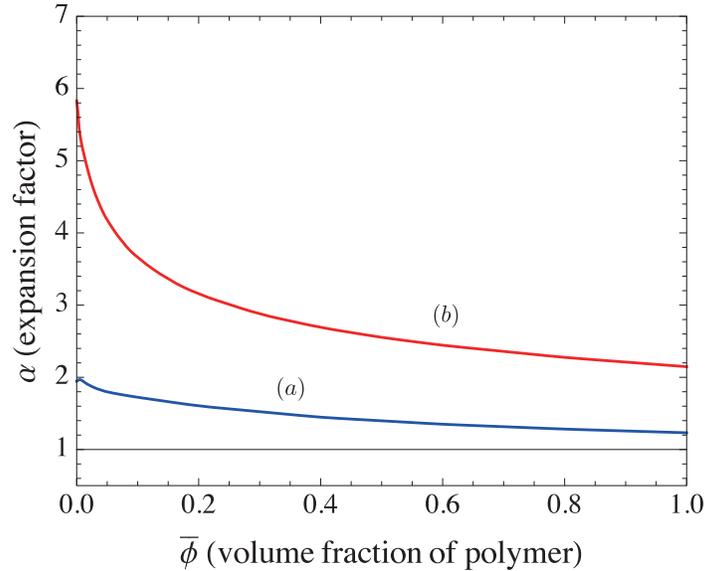}
\caption{The average volume fraction $\bar{\phi}$ dependence of the expansion factor, $\alpha$, for the hypothetical branched polymer system with (a) $N=10^{2}$ and (b) $N=10^{4}$ ($d=3$). Calculated numerically according to eq. (\ref{BEV-27}) with the help of eqs.(\ref{BEV-36}) and (\ref{BEV-37}) using the parameters shown in Table \ref{BPADC-Table}.}\label{Alpha-BP}
\end{center}
\end{figure}

The simulation results of eq. (\ref{BEV-27}) are summarized in Fig. \ref{Alpha-BP} for the polymers having (a) $N=10^{2}$ and (b) $N=10^{4}$ ($d=3$). It is seen that i) $\alpha$ decreases with increasing concentration, as expected, but ii) contrary to the case of linear molecules, the volume expansion still survives in the concentrated region; the molecules don't contract to the unperturbed size ($\alpha=1$) even in the melt state\cite{Wittmer8}.

Since, according to eqs. (\ref{BEV-19}) and (\ref{BEV-23}), the segment density inside a molecule increases as $\rho_{s}\propto N^{1-\nu d}$, we must have $\nu\ge 1/d$ in order for the density not to diverge. This is the packing density criterion that must be obeyed. The results, eqs. (\ref{BEV-31}) and (\ref{BEV-34}), for the dilution limit just satisfy this criterion, namely we observe that $\nu_{C\rightarrow 0}, \,\nu_{\Theta}\ge 1/d$. Our question is whether eq. (\ref{BEV-27}) satisfies $\nu\ge 1/d$ for all concentration range. In the following, we examine this problem.\\

We would like to emphasize that the curves shown in Figs. \ref{PSt-8}, \ref{PMMA-8} and \ref{Alpha-BP} do not necessarily represent unique solutions. From purely a mathematical point of view, it is impossible that such a complicated implicit function as eq. (\ref{BEV-27}) yields only one solution. In the simulations shown above, we have ignored physically unrealistic solutions. The excluded volume problem by no means belongs to an elementary science.

\subsection{Expansion Factor in Melt State}\label{EFMS}
\begin{figure}[H]
\begin{center}
\includegraphics[width=9.2cm]{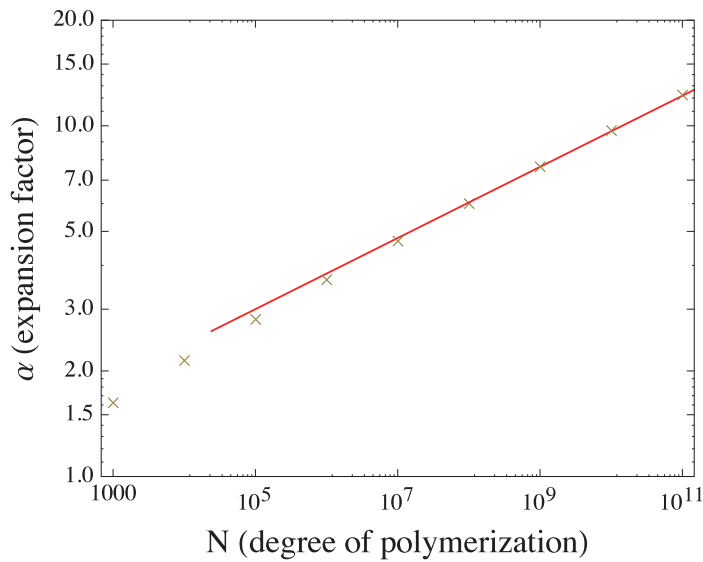}
\caption{$N$ dependence of the expansion factor in the melt state $\Bar{\phi}=1$: $\left(\times\right)$ simulation points according to eq. (\ref{BEV-27}) for $d=3$ with the help of eq. (\ref{BEV-37}) ($\omega=2$ and 3); the red solid line is the tangent, $\kappa=d\log \alpha/d\log N$, at $N=10^{11}$ having the value $\kappa\cong 1/10$.}\label{ExpandedCoil-B}
\end{center}
\end{figure}

\begin{figure}[h]
\begin{center}
\includegraphics[width=9.2cm]{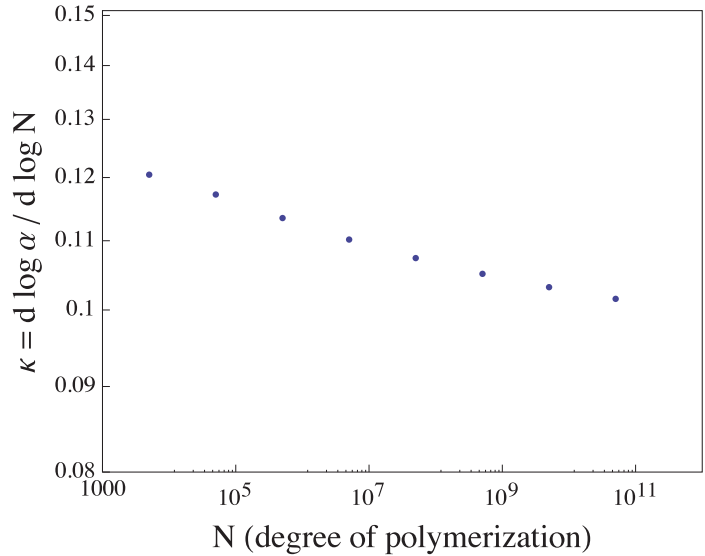}
\caption{$N$ dependence of the gradient, $\kappa=d\log \alpha/d\log N$, for the simulation points $\left(\times\right)$ in Fig. \ref{ExpandedCoil-B}. The gradient ($\kappa$) is not in the steady state, but falling slowly throughout this region, suggesting $\kappa\le\tfrac{1}{10}$.}\label{ExponentGRAD}
\end{center}
\end{figure}

The $N$ dependence of $\alpha$ in the melt state can be evaluated by putting $\Bar{\phi}=1$ in eq. (\ref{BEV-27}). We are interested in the exponent of the scaling hypothesis: $\alpha=\text{constant}\cdot N^{\kappa}$. The numerical results are illustrated with the symbols ($\times$) in Fig. \ref{ExpandedCoil-B}. It is seen that the gradient, $\kappa=d\log \alpha/d\log N$, is slowing down throughout the interval, $N=[10^{3}, 10^{11}]$; e.g., the numerical value at $N=10^{11}$ is $\kappa\approx 0.1$ (solid line), but still not in the steady state. The situation becomes clearer by inspecting the $\kappa$ vs $N$ plot shown in Fig. \ref{ExponentGRAD} which suggests strongly $\kappa\le 0.1$.

From this result we can evaluate the exponent $\nu$ defined by $\langle s_{N}^{2}\rangle^{1/2}\propto N^{\nu}\,\, (N\rightarrow\infty)$\cite{Stauffer8, Redner8, deGennes8, Issacson8, Seitz8, Parisi8, Daoud82, Lubensky8}. Since $\langle s_{N}^{2}\rangle^{1/2}=\alpha\,\langle s_{N}^{2}\rangle_{0}^{1/2}$ and $\langle s_{N}^{2}\rangle_{0}^{1/2}\propto N^{1/4}$\cite{Zim8, Dobson8, Kajiwara8} for randomly branched polymers, it follows that $\nu_{melt}\le\frac{1}{4}+0.1=0.35$. We have finally
\begin{equation}
0.33\dots\le\nu_{melt}\le 0.35\label{BEV-38}
\end{equation}
with the lowest bound $0.33\dots=1/3$ being the critical packing density ($\nu_{\text{critical}}=1/d$). The result is consistent with the value in the preceding paper\cite{Kazumi82} which was evaluated on the basis of the disappearance of the inhomogeneity.


\appendix
\section{More Problems}
Through the present work we have learned that a branched polymer undergoes the volume expansion under all conditions including the $\Theta$ state. The result raises a new question: Why is the theory of gelation not taking into consideration the excluded volume effects so successful? If a branched molecule expands, it may be expected that the end-to-end distance, $\langle r_{N}^{2}\rangle^{1/2}$, of a ``chain'' (embedded in the branched molecule) increases as well, which will lower the probability of cyclization. The total ring concentration, $[\Gamma(p)]$, in the R$-$A$_{f}$ model can be approximated in the form:
\begin{equation}
\left[\Gamma(p)\right]\simeq \sum_{k=1}^{\infty}\frac{\varphi_{k}}{2\NA\hspace{0.3mm}k}+(f-1)\sum_{k=1}^{\infty}\frac{\varphi_{k}}{2\NA}(p-p_{c_{0}})\hspace{7mm}\left(p_{c_{0}}\le p\le p_{c}\right)\label{BEV-39}
\end{equation}
with $\varphi_{k}$ being the relative cyclization frequency of a \textit{k}-chain, and $\NA$ the Avogadro number. It is important to notice that eq. (\ref{BEV-39}) is independent of the monomer concentration, $C_{0}$, in the system, the natural consequence of the invariance principle of ring concentration\cite{Kazumi81}. $\varphi_{k}$ has the form:
\begin{equation}
\varphi_x=\left(d/2\pi^{d/2}l_{s}^{\hspace{0.3mm}d}\right)\displaystyle\int_{0}^{d/2\nu_{\hspace{-0.3mm}x}}\hspace{-2mm}t^{\frac{d}{2}-1}e^{-t}dt\label{BEV-40}
\end{equation}
where $\nu_{x}=\langle r_{x}^{2}\rangle/l_{s}^{2}$, and $l_{s}$ the length of the cyclic bond. It is seen that the excluded volume effects have the influence, through the quantity $d/2\nu_{\hspace{-0.3mm}x}$, on the cyclization probability. So eq. (\ref{BEV-39}) has the weak concentration dependence in the gelling system. As  the volume expansion advances, the cyclization probability must decrease accordingly. As a result the assumption of the unperturbed chains tends to estimate excessively the ring concentration and hence overestimate the gel point, $p_{c}$, because of the equality: $p_{c}=p(\text{inter})+p(\text{ring})$.

The present work showed that $\alpha\hspace{-0.3mm}\approx\hspace{-0.3mm}\text{constant}\cdot N^{0.1}$ for a large $N$ in the melt state of the monodisperse system. Putting aside the problem of the applicability of the present theory to the gelling system with the broad dispersity, if we apply the present result to eq. (\ref{BEV-40}), the cyclization frequency, $\varphi_{k}$, falls to $\approx 70\%$ of the unperturbed chains ($\alpha=1$). This is not a minor change, but by no means a major one. It is most probable that the influence of the excluded volume effects on $p_{c}$ may be, in the event, absorbed into the uncertainty of the characteristic constant, $\langle\CF\rangle$\footnote{\, $\CF$ is a function of the chain length, $x$. It approaches a constant value as $x\rightarrow\infty$.}, along with the quality of the Gaussian approximation for short chains.
\vspace{2mm}


\end{document}